\documentclass[twocolumn, times, twocolappendix]{aastex701}
\usepackage{xspace}
\usepackage{amsmath}

\newcommand{\methanol}{CH$_3$OH\xspace}

\newcommand{\revise}[1]{\textrm{#1}}
\newcommand{\rerevise}[1]{\textrm{#1}}
\graphicspath{{./}{../figure/}}
\begin{document}

\title{Probing the Innermost Region of the V883 Ori Disk Using ALMA Band 1 Methanol Line Observations}

\author[0000-0003-4099-6941]{Yoshihide Yamato}
% \affiliation{Department of Astronomy, Graduate School of Science, The University of Tokyo, 7-3-1 Hongo, Bunkyo, Tokyo 113-0033, Japan}
\affiliation{RIKEN Pioneering Research Institute, 2-1 Hirosawa, Wako, Saitama 351-0198, Japan}
\email[show]{yyamato.as@gmail.com}

\author[0000-0003-2493-912X]{Shota Notsu}
\affiliation{Department of Earth and Planetary Science, Graduate School of Science, The University of Tokyo, 7-3-1 Hongo, Bunkyo-ku, Tokyo 113-0033, Japan}
\affiliation{RIKEN Pioneering Research Institute, 2-1 Hirosawa, Wako, Saitama 351-0198, Japan}
\email[]{shota.notsu@eps.s.u-tokyo.ac.jp}

\author[0009-0002-3884-5626]{Rui Zhuang}
\affiliation{Department of Astronomy, The University of Tokyo, 7-3-1 Hongo, Bunkyo, Tokyo 113-0033, Japan}
\affiliation{National Astronomical Observatory of Japan, 2-21-1 Osawa, Mitaka, Tokyo 181-8588, Japan}
\email[]{}

\author[0000-0003-3283-6884]{Yuri Aikawa}
\affiliation{Department of Astronomy, Graduate School of Science, The University of Tokyo, 7-3-1 Hongo, Bunkyo-ku, Tokyo 113-0033, Japan}
\email[]{}

\author[0000-0002-3297-4497]{Nami Sakai}
\affiliation{RIKEN Pioneering Research Institute, 2-1 Hirosawa, Wako, Saitama 351-0198, Japan}
\email[]{}

\begin{abstract}
The snowlines of major volatiles in protoplanetary disks play a pivotal role in dust evolutions and volatile delivery to nascent planetary systems. In this paper, we report the Atacama Large Millimeter/submillimeter Array Band 1 ($\approx7.5$\,mm) observations of methanol (\methanol) emission lines in the disk around the FU-Ori type star V883 Ori, where accretion outburst heats the disk and the majority of ices has sublimated. We detect three \methanol emission lines at an angular resolution of $\approx0\farcs2$. The stacked \methanol image exhibits a centrally-peaked morphology in contrast to the previous (sub-)mm observations that show a central depression. By fitting radially-resolved line profiles, we derive the radial intensity profile of the \methanol emission where we find a steep increase at $\lesssim40$\,au.
% while the same analysis of a Band 6 \methanol line shows no such increase.
The column density of \methanol reaches at least $\sim10^{19}$--$10^{20}$\,cm$^{-2}$ at $\sim10$\,au. This provides direct evidence that a significant amount of warm gaseous methanol is present in the innermost region of the disk where its emission has been suppressed in previous (sub-)mm observations due to the optically thick dust emission. 
% We provide an independent estimate of the methanol snowline radius of $\sim30$--55\,au, similar to the previous inference on the water snowline radius from the dust continuum observations. 
The steep increase in the intensity profile may indicate that the \methanol snowline in the midplane is located at $\sim30$--55\,au, or that the \methanol emission traces the temperature structure given that the emission is likely optically thick.
Our results demonstrate the capability and significance of (sub-)cm observations in probing the innermost opaque region of disks, paving the way for the future observations with upcoming facilities. 
% We suggest that the molecular-line-based snowline estimates with (sub-)mm observations are biased due to the optically thick dust, and they only trace the 

\end{abstract}

% \keywords{\uat{Galaxies}{573} --- \uat{Cosmology}{343} --- \uat{High Energy astrophysics}{739} --- \uat{Interstellar medium}{847} --- \uat{Stellar astronomy}{1583} --- \uat{Solar physics}{1476}}


\section{Introduction} \label{sec:intro}
The physical and chemical structures of protoplanetary disks plays a central role in shaping the initial conditions of planet formation and the delivery of volatiles to nascent planetary systems. 
% In particular, the inner a few to a few tens of au region of 
In particular, snowlines (sublimation fronts) of major volatiles, such as water and carbon monoxide, significantly affect the chemical composition in a disk by modifying the primary carriers of elements in both gas and solid phases \citep[e.g.,][]{Oberg2011}. The snowline of water is located at $\sim1$--5\,au in a disk around a typical solar-mass pre-main sequence star, and drastically changes the elemental abundance ratios of gas and ice as well as the properties of dust grains due to the desorption of majority of ices, which makes it crucial in understanding the dust evolution and disk chemistry \citep[e.g.,][]{Oka2011, Eistrup2016, Banzatti2023, Aikawa2024}. 

% Water snowline depends on the disk temperature structure and is typically located at $\sim$1--5\,au in the disk midplane around a solar-type star ($L_\star\approx1\,L_\odot$) in Class I--II phases. While warm water vapor emission, along with complex organic molecules (COMs) that have a similar sublimation temperature ($\sim100$\,K), has frequently been detected in younger Class 0 sources as they have warmer condition \citep{}, observations of water and COMs in Class II disks are challenging due to their small spatial scales of water sublimation region. 

% Recent sensitive observations with the Atacama Large Millimeter/submillimeter Array (ALMA) have enabled to detect COM emission in warmer disks around Herbig Ae/Be stars, but they lack detection of water 

% While spatially resolving the water snowline with molecular line observations are typically infeasible with current facilities, 
Accretion outburst provides a unique opportunity to probe the water snowline and the sublimated ices interior to it, as the disk temperature increases due to the outburst and the water snowline has shifted outward. V883 Ori, a low-mass ($M_\star\approx1.3\,M_\odot$; \citealt{Cieza2016}) FU Orionis type object located in the Orion nebula cluster ($d\sim400$\,pc; \citealt{Strom1993}), is one of such sources with an elevated bolometric luminosity of $\sim200\,L_\odot$ \citep{Furlan2016}. It is in a transition phase between Class I and Class II, and harbors a large ($\sim300$\,au), massive ($\sim0.3M_\odot$) Keplerian-rotating disk \citep{Cieza2016}, which has been frequently observed mainly in (sub-)mm wavelengths using the Atacama Large Millimeter/submillimeter Array (ALMA). \citet{Cieza2016} first presented a high-resolution ($\approx0\farcs03$ or 12 au) dust continuum image of the disk at 1.3\,mm, and identified an intensity break and an adrupt change in optical depth at 42\,au. These were interpreted as an outcome of the ice sublimation at water snowline that changes the optical property of dust grains. \citet{vantHoff2018} and \citet{Lee2019} detected abundant warm gas of \methanol and other complex organics with ALMA Band 7 observations, making this source an ideal target to study the ice composition in protoplanetary disks. Following these studies, extensive observations of molecular emission lines, including those of complex organics and water, have been performed in (sub-)mm wavelengths at moderate- to high-resolutions ($\sim0\farcs1$--0\farcs5) \citep[e.g.,][]{Tobin2023, Yamato2024_V883Ori, Lee2024, Fadul2024b, Fadul2024a, Jeong2025, Leemker2025, Zeng2025, Nakasone2026}. For instance, \citet{Tobin2023} detected an H$_2$$^{18}$O line in Band 5 and an HDO line in Band 6 at $\approx0\farcs1$ resolution, and argue that the water snowline is located at $\sim80$\,au, which is larger than the estimate based on the continuum emission \citep{Cieza2016}. 
% \citet{Leemker2021} also suggested a larger snowline radius of 75--120\,au based on the distributions of HCO$^+$ that is expected to be chemically anti-correlated with water. 
% Recently, \citet{Nakasone2026} reported the detections of H$_2$$^{18}$O and HDO line emission at ALMA Band 7, and found that the Band 7 HDO line is significantly
% weaker than expected based on the rotational diagram derived from the Band 6 HDO lines reported by \citet{Tobin2023}. 
% They discussed that considering the higher upper-state energies ($>300$ K) of the Band 7 lines, this attenuation can be explained by a more compact, hotter emitting region with an effective radius of $\sim53$ au and/or frequency-dependent dust absorption that enlarges the apparent inner cavity at a higher frequency.

However, molecular line observations in (sub-)mm wavelength suffer from the effect of the optically thick dust emission. The distributions of water and complex organic emission in ALMA Band 6/Band 7 high-resolution ($\approx0\farcs1$) images extend out to $\sim160$\,au while commonly shows an inner emission cavity at $\lesssim40$\,au \citep{Lee2019, Tobin2023, Jeong2025}. The radius of the emission cavity well corresponds to the intensity break seen in the continuum emission \citep{Cieza2016, Houge2024}, suggesting that the molecular line emission is suppressed by optically thick dust emission. This makes it challenging to probe the innermost $<40$\,au radius with the molecular line observations.
% and test the continuum-based prediction on the snowline radius of 42\,au with molecular line observations. 
% \citet{vantHoff2018} suggested that the outer extent of the \methanol emission could trace either inside the water snowline or the snow surface beyond the midplane snowline, depending on the disk temperature structure. 
In light of this, \citet{Yamato2024_V883Ori} performed ALMA observations using Band 3 ($\approx3$\,mm) where dust optical depth expected to be smaller, but the radial intensity profiles of complex organic lines similarly show a depression at $\lesssim40$\,au, suggesting that the dust continuum is still optically thick, or that molecules might be efficiently destroyed by strong UV and/or X-ray radiation from the central eruptive star \citep[e.g.,][]{Notsu2021}. To distinguish these scenarios, further longer-wavelength observations are needed. 

In this paper, we present ALMA Band 1 ($\approx7.5$\,mm) observations of V883 Ori at an angular resolution of $\approx0\farcs2$ targeting multiple \methanol lines. We describe the summary of the Band 1 observations as well as the complimentary archival Band 6 observations, and the analysis results in Section \ref{sec:observation}. 
% We present the spatial distributions and line profiles of the \methanol emission in Section \ref{sec:analysis}, along with the line profile analysis to derive radial profiles of intensity and column density. 
In Section \ref{sec:discussion}, we discuss the \methanol distributions and its implications for the thermal structure of the V883 Ori disk. We finally summarize our conclusion in Section \ref{sec:summary}.

% In particular, the inner regions of disks, within several tens of astronomical units from the central protostar, are the sites where terrestrial planets and the cores of giant planets are expected to form. Characterizing the molecular inventory of these regions is therefore essential for understanding how chemical complexity is established and evolves during the early stages of star and planet formation.

% In recent years, high spatial resolution observations at (sub-)millimeter wavelengths have been powerful in characterizing the physical and chemical structures of disks. Both dust continuum and molecular line emission can be observed in the outer region of disks (a few tens to hundreds of au), providing fruitful insight into the disk substructure, thermal structure, and chemical composition. Probing the chemistry in the inner disk, however, has proven to be challenging. At (sub-)millimeter wavelengths, molecular line observations are often hampered by high dust optical depths due to large column densities in the inner region. 

\section{\rerevise{Observations}} \label{sec:observation}
% \subsection{ALMA Observations}
We observed the V883 Ori disk using Band 1 during ALMA Cycle~11, targeting multiple \methanol lines. In addition, we complimentarily used the archival data of \methanol lines in Band 6 originally published in \citet{Tobin2023}. Details of the observations and data reduction can be found in Appendix \ref{appendix:observations}. Briefly, we produced image cubes of three detected \methanol lines in Band 1 with a channel width of $\approx1.0$\,km\,s$^{-1}$. The spectroscopic properties, beam sizes, and noise levels are summarized in Table \ref{tab:lines}. We also produced an aggregate continuum image at $\approx7.5$\,mm using line-free channels of the data. For Band 6 \methanol lines, we specifically focused on the unblended line at 241.852\,GHz ($\nu_\mathrm{t}=0$, $J_{K_a,K_c}=5_{3,2}$--$4_{3,1}$, $E_\mathrm{u}=97.5$\,K) with original spatial and velocity resolutions of $\approx0\farcs1$ and 0.05\,km\,s$^{-1}$, respectively. We aligned the beam sizes and velocity channel samplings of all image cubes, and then used them in the subsequent analysis.

\begin{deluxetable}{lcCccCc}
\label{tab:lines}
\tablecaption{Observed \methanol Lines}
\tablehead{\colhead{Transition} & \colhead{$\nu_0$} & \colhead{$\log_{10}A_\mathrm{ul}$} & \colhead{$g_\mathrm{u}$} & \colhead{$E_\mathrm{u}$} & \colhead{Beam (P.A.)} & \colhead{RMS}\\ \colhead{ } & \colhead{($\mathrm{GHz}$)} & \colhead{(s$^{-1}$)} & \colhead{ } & \colhead{($\mathrm{K}$)} & \colhead{ } & \colhead{($\mathrm{K}$)}}
\startdata
CH$_3$OH $\nu_\mathrm{t}=0$ $J_{K_a,K_c} =$ $6_{2, 5}$--$5_{3, 2}$ & 38.293270 & -7.3218 & 52 & 86.5 & 0\farcs21\times0\farcs14\,(61\arcdeg) & 15 \\
CH$_3$OH $\nu_\mathrm{t}=0$ $J_{K_a,K_c} =$ $6_{2, 4}$--$5_{3, 3}$ & 38.452629 & -7.3161 & 52 & 86.5 & 0\farcs21\times0\farcs14\,(62\arcdeg) & 14 \\
CH$_3$OH $\nu_\mathrm{t}=0$ $J_{K_a,K_c} =$ $7_{0, 7}$--$6_{1, 6}$ & 44.069366 & -6.3897 & 60 & 65.0 & 0\farcs18\times0\farcs12\,(60\arcdeg) & 20 \\
% CH$_3$OH $v=0$ $J_{K_a,K_c} =$ $10_{1, 9}$--$10_{1, 10}$ & 45.818115 & -8.1055 & 84 & 143.3 & 0\farcs48\times0\farcs16\,(-85\arcdeg) & 15 \\
% CH$_3$OH $v=1$ $J_{K_a,K_c} =$ $9_{3, 7}$--$10_{2, 9}$ & 45.843555 & -6.8053 & 76 & 152.2 & 0\farcs48\times0\farcs16\,(-85\arcdeg) & 14
\enddata
\tablecomments{The spectroscopic properties of the lines are taken from the Cologne Database for Molecular Spectroscopy (CDMS; \citealt{CDMS1, CDMS2, CDMS3}) with the original data from \citet{Lees1968}.}
\end{deluxetable}

\section{\rerevise{Analysis \& Results}} \label{sec:analysis}
\subsection{Spatial Distributions and Line Profiles}\label{subsec:distributions}
\begin{figure*}
\epsscale{1.1}
\plotone{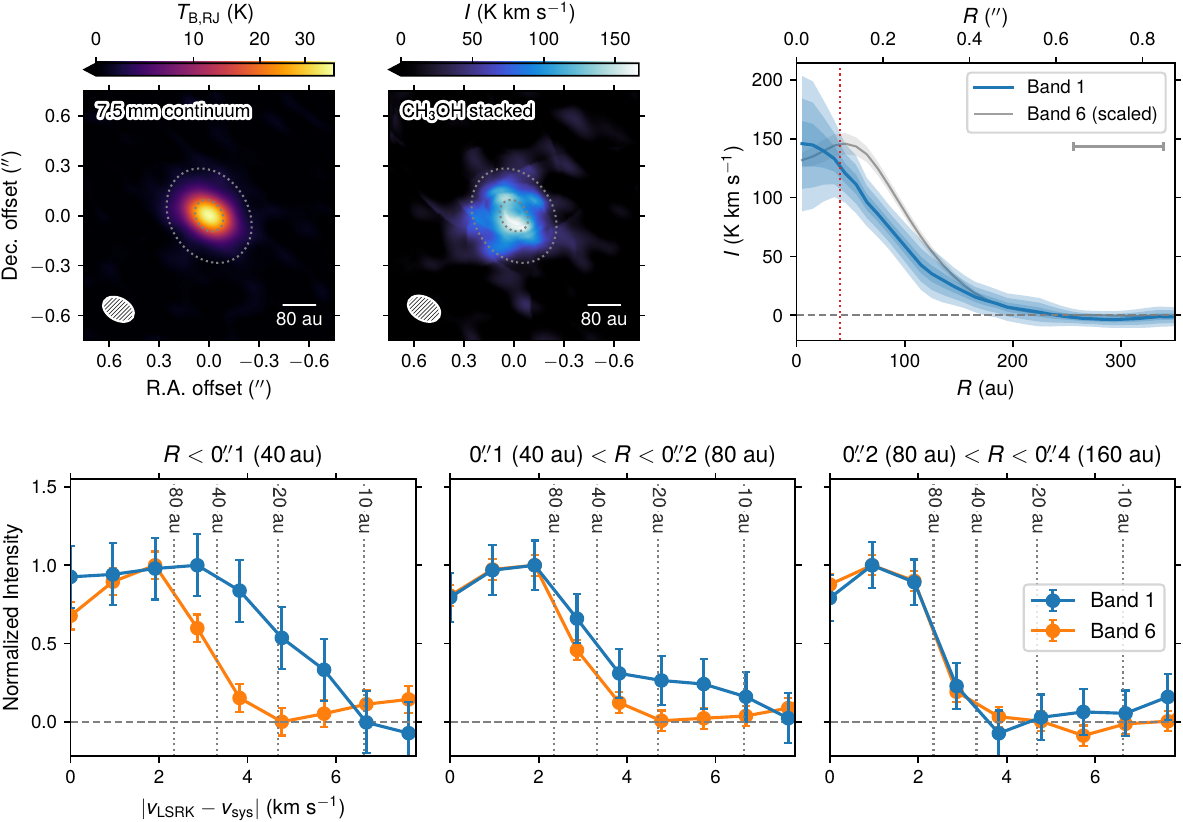}
\caption{Top row: 40\,GHz ($\approx7.5$\,mm) continuum image (left), Band 1 stacked CH$_3$OH zeroth moment map (middle), and its deprojected radial intensity profile (right) in the V883 Ori disk. The synthesized beam and a scale bar are shown in the lower left and right corner, respectively, for the left and middle panels. For a visual guide, the gray dotted ellipses indicate 40 au and 120 au radii, roughly delineating the sharp edges of the inner and outer dust disks, respectively, seen in the higher-resolution continuum images (see \citealt{Cieza2016}). In the right panel, the radial intensity profile of a CH$_3$OH line in Band 6 is shown in gray for comparison. The Band 6 profile are scaled so that the radial peak value becomes the same as that of the Band 1 profile. The shaded regions indicate [1, 2, 3]$\sigma$ uncertainties for the Band 1 profile, while only 1$\sigma$ uncertainty is shown for the Band 6 profile for visual clarity. The vertical red dotted line marks 40 au radius. The FWHM of the beam major axis is shown in the upper right corner. Bottom row: Comparison of the folded line profiles extracted from different annuli, $R<0\farcs1$ (left), $0\farcs1<R<0\farcs2$ (middle) and $0\farcs2<R<0\farcs4$ (right). The profiles for both Band 1 (stacked; blue) and Band 6 (orange) CH$_3$OH lines are shown. The error bars indicate $1\sigma$ uncertainty. The vertical dotted lines mark the Keplerian velocities corresponding to 80\,au, 40 au, 20 au, and 10 au.}
\label{fig:mom0_radial_profile}
\end{figure*}

We present the continuum emission at $\approx7.5$\,mm, the velocity-integrated intensity (zeroth moment) map stacked for three \methanol lines, and its deprojected radial intensity profile in Figure \ref{fig:mom0_radial_profile}. The zeroth moment map for each line was produced using the Python package \texttt{bettermoments} \citep{bettermoments} with a Keplerian mask generated by the Python script \texttt{keplerian\_mask.py} with known properties of the V883 Ori disk: $\mathrm{P.A.}=32\arcdeg$, $i=38.3\arcdeg$, $M_\star=1.29\,M_\odot$, and $v_\mathrm{sys}=4.3$\, km s$^{-1}$. \revise{The maps for individual transitions are shown in Figure \ref{fig:mom0_individual} in Appendix \ref{appendix:individual_mom0} along with potential caveats in interpreting the stacked map.} These maps were then stacked for all three \methanol lines by taking the mean of each pixel. We computed the radial intensity profiles by deprojecting and azimuthally averaging the stacked zeroth moment map, where we assumed a flat emission surface for deprojection. Both map and radial intensity profile are marginally spatially resolved and appears as approximately centrally peaked, suggesting that the line emission is originating from the inner 40\,au in addition to the outer region. For comparison, we show in Figure \ref{fig:mom0_radial_profile} the deprojected radial intensity profiles of the \methanol line in Band 6 that is produced in the same manner as the Band 1 \methanol profile. The Band 6 profile shows slight dip at $\lesssim40$\,au radius, corresponding to the 40\,au inner emission hole as seen in the higher-resolution images \citep[e.g.,][]{Lee2019, Tobin2023, Lee2023}. 

Given that zeroth moment map and its deprojection suffer from the beam convolution effect and could smear out any intrinsic structures, we also examine line profile that hold the information of the emitting radii under the assumption of gas kinematics. 
% Assuming the emitting gas follows the Keplerian rotation, the presence of high-velocity components in line profiles indicate the emission from the inner region. 
We compute the line profile of individual Band 1 \methanol lines averaged over different deprojected annuli and take weighted average of them using a weight of $I_\mathrm{peak}/\sigma^2$, where $I_\mathrm{peak}$ is the maximum value of the individual line profiles and $\sigma$ is the RMS noise level measured on the line-free region. To improve S/N, we further stacked the red- and blue-shifted sides by folding the stacked line profile with respect to $v_\mathrm{sys}$ ($\approx4.3$\,km\,s$^{-1}$), assuming a symmetric line profile, characteristic to Keplerian rotation. Practically we regrid the velocity axis so that the profiles have symmetric velocity sampling with respect to $v_\mathrm{sys}$, and then both sides are averaged with equal weights. We also performed the same procedures for the Band 6 single \methanol line. The bottom panels of Figure \ref{fig:mom0_radial_profile} compare the stacked-and-folded line profiles of the Band 1 lines and Band 6 line at different radial regions. The line profiles of the Band 1 \methanol line at $R<0\farcs1$ clearly shows an excess at $\sim3$--6 km s$^{-1}$ region compared to the Band 6 profile, while they appear more similar in the outer region ($R>40$\,au). This is clear evidence that the Band 1 \methanol emission traces more inner region where Keplerian velocity is larger, in addition to the outer region.  

\subsection{Radially-resolved Line Profile Fit}\label{subsec:line_profile_fit}
\begin{figure*}
\epsscale{1.15}
\plotone{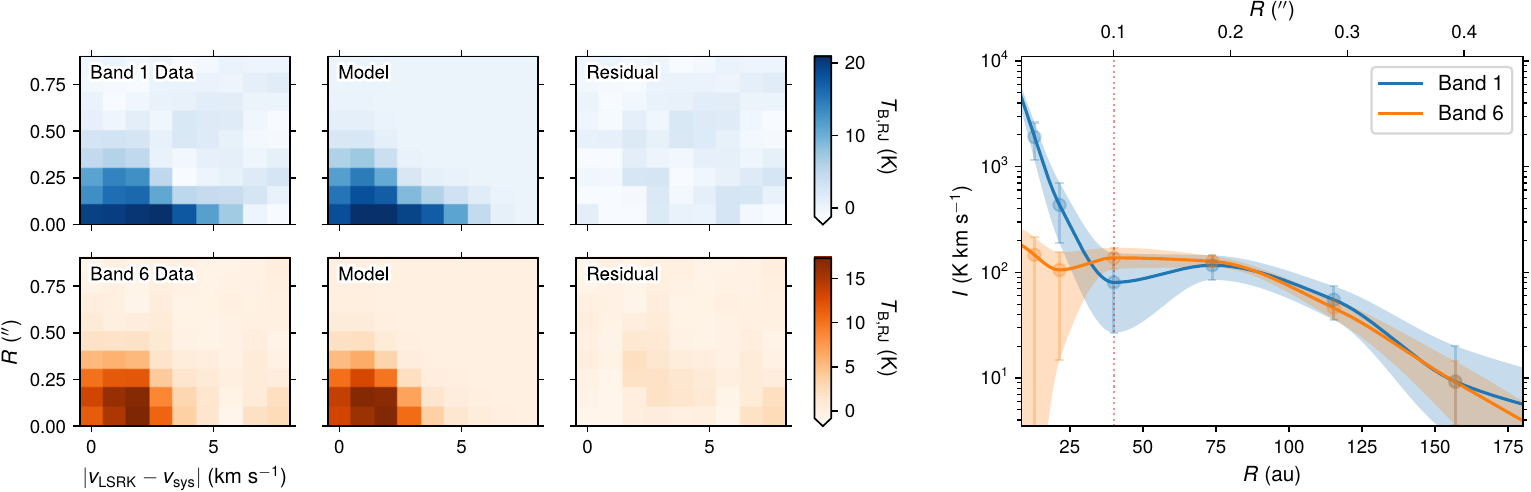}
\caption{Left: Radially-resolved, folded line profiles (left column), best-fit models of the fits (middle column), and their residuals (right column) for Band 1 (top row) and Band 6 (bottom row) \methanol data. \rerevise{The $1\sigma$ uncertainties of the radially resolved, folded line profiles vary with radius, increasing toward smaller radii, and range from 0.80 to 4.1 K for Band 1 and from 0.33 to 1.5 K for Band 6.} Right: Reconstructed radial intensity profiles from the line profiles fits for Band 1 (blue) and Band 6 (orange) \methanol lines. The fitted radial grid points are shown in transparent circles and associated error bars, while we also plot the piecewise cubic Hermite interpolation of these data points and their uncertainties by solid lines and color-shaded regions for visualization purpose. The data points and uncertainties are medians and 68\% highest density interval of the MCMC posterior distributions, respectively. The vertical red dotted line indicates 40 au radius. 
% The gray dashed curve shows the high-resolution ($\approx0\farcs1$) radial intensity profile of the Band 6 \methanol line extracted from the velocity-integrated intensity map \citep{Tobin2023}.
}
\label{fig:line_profile_fit}
\end{figure*}

To quantify the emission excess and convert it to the radial distributions, we employed a forward modeling approach to fit the line profiles at different annuli, as detailed in Appendix \ref{appendix:modeling}. Briefly, we numerically compute the model line profile and fit it to the observed line profiles extracted from multiple annuli. We prepared the radially-resolved line profiles (or ``R-V map'') by taking azimuthal averaging of the (stacked) \methanol line cubes within the deprojected concentric annuli with a width of $0\farcs1$ and folding with respect to $v_\mathrm{sys}$ (see Section \ref{subsec:distributions}) as shown in Figure \ref{fig:line_profile_fit}.
% While this method is in principle similar to the ones used in previous studies \citep{Bosman2021, Facchini2024, Yamato2024_V883Ori}, it takes the beam smearing effect into account most accurately. 
We assume that the emission is originating from the midplane of an axisymmetric disk with a radial intensity profile $I(r)$, and compute the model cube as
\begin{equation}\label{eq:model_cube}
    I_\nu(v; r, \phi) = \frac{I(r)}{\sqrt{2\pi}\sigma_v}\exp\left[-\frac{(v - v_\mathrm{LOS}(r, \phi))^2}{2\sigma_v^2}\right],
\end{equation}
where we assume that the local line shape is Gaussian with an intrinsic line width of $\sigma_v = 0.25$\,km\,s$^{-1}$ (comparable to the thermal line width of \methanol at $T=200$\,K), and $v_\mathrm{LOS}$ is the line-of-sight velocity at a disk polar coordinate $(r,\phi)$;
\begin{equation}\label{eq:los_velocity}
    v_\mathrm{LOS}(r,\phi) = \sqrt{\frac{GM_\star}{r}}\cos\phi\sin i + v_\mathrm{sys},
\end{equation}
where $G$ is the gravitational constant. After taking the effect of beam convolution into account (see Appendix \ref{appendix:modeling} for a detailed formulation of this) and taking azimuthal averaging at each deprojected annulus, we fit the model R-V map to the observed one to estimate $I(r)$. We employ a Bayesian inference and practically use the affine-invariant Markov Chain Monte Carlo (MCMC) algorithm implemented in the Python package \texttt{emcee} \citep{emcee} to sample the posterior distributions of $I(r)$. The details can be found in Appendix \ref{appendix:modeling}.  

Figure \ref{fig:line_profile_fit} shows the comparison of the data and the best-fit models that maximize the likelihood, and the reconstructed radial intensity profiles of the Band 1 and Band 6 \methanol lines. The best-fit models that maximize the likelihoods reproduce the data well, including the excess of the Band 1 line profile at $|v_\mathrm{LSRK}-v_\mathrm{sys}|\gtrsim3$\,km\,s$^{-1}$. While the reconstructed intensity profiles are similar between Band 1 and Band 6 lines in the outer region ($>40$\,au), the Band 1 profile shows a steep increase at $\lesssim40$\,au where the Band 6 profile are flat. The Band 1 intensity is higher than the Band 6 one by a factor of $\gtrsim10$ in the innermost region ($\lesssim20$\,au). \revise{A comparison of the radial intensity profiles derived from line profile fit and image deprojection can be found in Appendix \ref{appendix:profile_comparison}.}
% The derived intensity profile of the Band 6 \methanol line is slightly offset compared to the high-resolution ($\approx$0\farcs1) radial intensity profile directly extracted from the (beam-convolved) integrated intensity map as shown in gray dashed curve, which is consistent each other considering the beam dilution effect. \revise{We note that the reconstructed radial intensity profile of a \methanol transition in Band 3 also show a similar distribution as the Band 6 profiles \citep{Yamato2024_V883Ori}.}

% \begin{figure*}
% \epsscale{1.15}
% \plotone{V883Ori_line_profile_comparison.pdf}
% \caption{Comparison of the line profiles extracted from different annuli, $R<0\farcs1$ (left), $0\farcs1<R<0\farcs2$ (middle) and $0\farcs2<R<0\farcs4$ (right). The profiles for both Band 1 (stacked; blue) and Band 6 (orange) CH$_3$OH lines are shown. The profiles are folded at the systemic velocity and the red- and blue-shifted sides are stacked assuming a symmetric line profile (see text for details). The vertical dotted lines mark the Keplerian velocities corresponding to 80\,au, 40 au, 20 au, and 10 au.
% % While the profiles at larger radii looks similar between Band 1 and Band 6, a significant excess ($\gtrsim3\sigma$) at high-velocities can be found in the Band 1 profile in the innermost radii.
% }
% \label{fig:line_profile_comparison}
% \end{figure*}

\subsection{\methanol Column Density and Line Optical Depth}
We convert the reconstructed radial intensity profile of the (stacked) Band 1 \methanol lines to the column density profile assuming (1) an optically thin emission, (2) a local thermodynamical equilibrium (LTE) condition, and (3) a fixed gas temperature $T$. We take into account the fact that the intensity profile is the weighted-average of three \methanol transitions with different intrinsic strengths as follows,
\begin{equation}
    N(r) = \left.I(r)\sum_iw_i\middle/\sum_i\frac{w_ig_{\mathrm{u},i}e^{-E_{\mathrm{u},i}/k_\mathrm{B}T}}{\gamma_{\mathrm{u},i}Q(T)}\right.,
\end{equation}
where $w_i$ is the averaged weights used for line stacking, $g_{\mathrm{u}, i}$ is the statistical weight of the upper state, $E_{\mathrm{u},i}$ is the upper state energy, $k_\mathrm{B}$ is the Boltzmann constant, $Q(T)$ is the partition function of \methanol at a temperature $T$, and $\gamma_{\mathrm{u},i} = 8\pi k_\mathrm{B}\nu^2/A_{\mathrm{ul}, i}hc^3$ where $\nu_0$ is the observing frequency of each transition, $A_{\mathrm{ul},i}$ is the Einstein A coefficient, $h$ is the Planck constant, and $c$ is the speed of light. The partition function is taken from the Cologne Database for Molecular Spectroscopy \citep[CDMS][]{CDMS1, CDMS2, CDMS3}. For the fixed gas temperature, we used the previous inferences based on the dust continuum and molecular line observations. \textrm{At $< 40$\,au, we used the radial dust temperature profile from a multi-wavelength analysis of the continuum emission at $\sim0\farcs1$ resolution (Zhuang et al. submitted), assuming that dust and gas are well mixed and share the same temperature. The temperature ranges from $\sim300$\,K and $\sim120$\,K between 10\,au and 40\,au.} We assumed a constant temperature of 120\,K at $> 40$\,au, which smoothly connect to the temperature at $<40$\,au, based on the previous COM line observations \citep{vantHoff2018, Lee2019, Yamato2024_V883Ori, Jeong2025}. 

\begin{figure}
% \epsscale{1.15}
\plotone{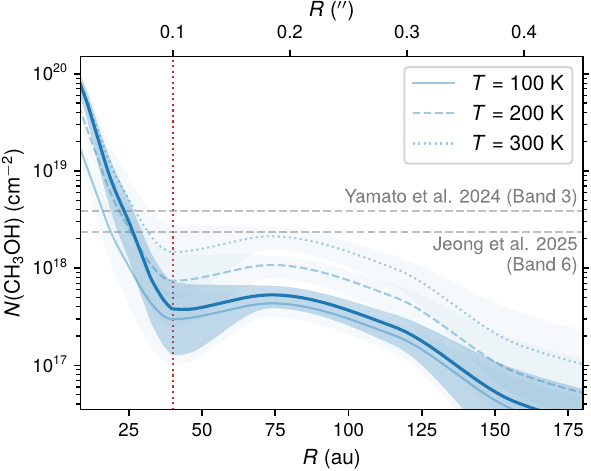}
\caption{\methanol column density profiles calculated from the \rerevise{reconstructed} Band 1 radial intensity profile. The thick blue line shows the profiles derived using the dust temperature profile (see text). The thin lines with different line styles indicate the profiles derived using constant temperatures of 100, 200, and 300\,K. The disk-averaged column densities from the previous studies using Band 3 and Band 6 data \citep{Yamato2024_V883Ori, Jeong2025} are marked by the horizontal gray dashed lines for comparison. The vertical red dotted line indicates 40 au radius.}
\label{fig:column_density_profile}
\end{figure}

Figure \ref{fig:column_density_profile} shows the resulting column density profile. For reference, we also show the column density profiles when assuming $T=100, 200,$ and 300\,K across the entire disk. The column density reaches $\gtrsim10^{19}$\,cm$^{-2}$ in the innermost region ($\lesssim20$\,au), which is higher than the previous disk-averaged estimates using Band 3 and Band 6 data \citep{Yamato2024_V883Ori, Jeong2025}. Note that the column densities here are lower limits as we assumed an optically thin emission. Indeed, the column density in the outer region ($\gtrsim40$\,au) are lower than the previous Band 3/Band 6 estimates using only optically thin lines and a gas temperature of 100--120\,K, suggesting that the Band 1 \methanol lines are also optically thick and thus column densities are underestimated. 
% As for the inner region ($\lesssim40$\,au),  

% Figure \ref{fig:column_density_profile} shows the resulting column density profiles for different assumed gas temperature of 100, 200, and 300\,K. These assumed temperatures are based on the previous estimates from complex organic and water lines (100--120\,K for COMs and 200 K for water; \citealt{Yamato2024_V883Ori, Jeong2025, Tobin2023, Nakasone2026}) except for 300 K, which likely corresponds to the temperature in the innermost region ($\lesssim20$\,au; e.g., \citealt{Cieza2016}). The column density reaches $\gtrsim10^{19}$\,cm$^{-2}$ in the innermost region, which is higher than the previous disk-averaged estimates using Band 3 and Band 6 data \citep{Yamato2024_V883Ori, Jeong2025}. Note that the column densities here are lower limits as we assumed an optically thin emission. Indeed, the column density in the outer region ($\gtrsim40$\,au) are lower than the previous Band 3/6 estimates using only optically thin lines and a gas temperature of 100--120\,K, suggesting that the \methanol lines in Band 1 are also optically thick and thus column densities are underestimated. 

\section{Discussion} \label{sec:discussion}
% \subsection{Warm Gaseous \methanol at $R<40$\,au}
The \methanol emission in the inner region ($\lesssim40$\,au) is direct evidence of the presence of the warm \methanol gas there, which has been hidden from view in the higher-frequency observations (ALMA Bands 3--7). At (sub-)mm wavelengths, the molecular line emission from there is obscured by the optically thick dust emission as suggested by previous studies \citep{Lee2019, Tobin2023, Yamato2024_V883Ori, Jeong2025}. 
% \citet{Yamato2024_V883Ori} conducted the first molecular line observations in ALMA Band 3 ($\approx3$\,mm), but concluded that the molecular line emission are still absent at $\lesssim40$\,au region possibly because the dust optical depth is still not lowered enough for the molecular line emission to be detected, or the gas-phase molecules there are destroyed by the strong incident UV and/or X-ray. 
Even in Band 3 ($\approx3$\,mm), the \methanol emission exhibits a central depression, from which \citet{Yamato2024_V883Ori} suggest that the thermal dust emission is still optically thick, or molecules might be destroyed by the strong incident UV and/or X-ray \citep[e.g.,][]{Notsu2021}.
\revise{Although the destruction of molecules may partly contribute to determining the gas-phase abundance, particularly in the innermost regions exposed to strong UV and X-ray radiation,}
the present Band 1 observations ($\approx7.5$\,mm) indicate that the absence of molecular line emission at $\lesssim40$\,au in higher-frequency observations can indeed be \revise{largely} explained by the dust optical depth, which is likely lowered enough at sub-cm wavelength. \textrm{This is also consistent with recent study that argues the thermal dust emission is optically thin ($\tau\lesssim 0.1$ at $\lesssim40$\,au) and dominated by the free-free emission in the VLA Q Band observations ($\approx7$\,mm; Zhuang et al. submitted). Based on the multi-band continuum observations, Zhuang et al. (submitted) also suggests that the dust absorption optical depth is slightly lower than unity in Band 3 ($\approx3.1$\,mm), although the exact values depends on the assumed dust model. This seemingly contradicts with the central depression seen in the molecular line observations in Band 3 \citep{Yamato2024_V883Ori}. However, as the appearance of the molecular line emission also depends on the vertical stratifications of gas and dust and the dust scattering properties \citep[e.g.,][]{Bosman2021}, a more sophisticated, combined analysis of dust continuum and molecular line emission is required to infer the relation between dust and molecular line emission.} In any case, our results suggest that (sub-)cm observations are crucial in probing the molecular gas in the opaque inner region, as also demonstrated for the younger Class 0 source NGC1333~IRAS4A1 \citep{DeSimone2020}. 
% In Zhuang et al. submitted, the dust optical depth at $\lesssim40$\,au is greater than unity 
% \todo{to be confirmed with Ohashi-san}). 

% Our analysis suggests that the \methanol emission is present at least down to $\sim10$\,au, indicating that the destruction of molecules are not efficient at $\gtrsim$10\,au. \todo{can be compared with models more quantitatively?}

% The appearance of the molecular line emission also depends on the vertical distributions of gas and dust and the dust albedo (i.e., scattering efficiency), in addition to the dust (absorption) optical depth (e.g., see \citealt{Bosman2021}).

The \methanol column density profile derived in the present work also have significant implications for the snowline radius and the thermal structure of the V883 Ori disk. Previous estimates on the water snowline radius range between $\approx40$--120\,au, depending on tracers \citep{Cieza2016, vantHoff2018, Lee2019, Leemker2021, Tobin2023}. The \methanol column density profile based on Band 1 observations suggest that the \methanol column density steeply increase at $\sim40$\,au (Figure \ref{fig:column_density_profile}). 
\revise{Although the column density profile may partly reflect the underlying gas surface density structure, we would expect a steep increase in column density at the snowline, caused by the drastic enhancement of the gas-phase abundance due to ice sublimation.}
We therefore suggest that the \methanol sublimation radius at the disk midplane is located at $\sim30$--55 au\footnote{We determine the range based on the radial grid spacing (see Figure \ref{fig:line_profile_fit}).}, while the emission in the outer region ($\gtrsim40$\,au) is originated from the upper layer of the disk (i.e., snow surface). \citet{vantHoff2018} argue that the Band 6 \methanol lines can trace either the snowline in the midplane or the snow surface in the disk upper layer depending on the vertical temperature structures, where our results support the latter scenario. An empirical estimate on the emission surface of HDO and \methanol lines in Band 6 also suggests elevated emission surfaces ($z/r \approx0.4$--0.6) in the outer region ($\gtrsim40$\,au; \citealt{Leemker2024_thesis}). 
% \todo{We need to show abundance (not column density) jump at 40 au to better argue the snowline radius}

Alternatively, it is also possible that the \methanol intensity profile traces the temperature distribution rather than the column density if the emission is optically thick across the entire region. Indeed, the column density derived under the assumption of an optically thin emission is lower than the previous multi-line estimates in the outer region ($\gtrsim40$\,au; \citealt{Jeong2025, Yamato2024_V883Ori}), suggesting that the line emission at $\gtrsim40$\,au is optically thick. The inner region is also expected to be optically thick in the \methanol emission due to the high column density, although the line optical depth also depends on the gas temperature. \textrm{To calculate the column density we assume that the gas temperature is the same as the dust temperature derived from the multi-wavelength continuum observations (Zhuang et al. submitted), but the actual emitting heights of dust and gas (and thus temperatures) can be different.
% , particularly in the outbursting source where the disk midplane can be heated by the viscous accretion heating. 
Future multi-line, higher sensitivity observational efforts would be needed to independently estimate the gas temperature and column densities, and in turn, to determine the radial/vertical thermal structure of the disk.}  

If the steep increase in the \methanol intensity profile at $\sim40\,$au indeed corresponds to the \methanol sublimation radius in the midplane, the water snowline should also be located at a similar or slightly smaller radius given that the binding energy of water is similar or slightly larger than that of \methanol \citep[][and references therein]{Minissale2022}. This supports the previous inference of 42 au snowline radius based on the continuum emission \citep{Cieza2016}, suggesting that the change in the dust opacity is due to the snowline effects, i.e., a dust pileup interior to the snowline \citep{Birnstiel2010, Banzatti2015, Pinilla2016} or water evaporation and re-coagulation of silicates \citep{Schoonenberg2017}. \textrm{Zhuang et al. (submitted) also suggest that the dust grain size may drastically change at $\sim40$\,au ($\gtrsim1$\,cm at $>40$\,au and $\lesssim1$\,mm at $<40$\,au) based on the multi-wavelength analysis of the high-resolution continuum observations.} In contrast, previous molecular-line-based estimates in (sub-)mm wavelengths of $\approx75$--120\,au \citep{Lee2019, Leemker2021, Tobin2023} are inconsistent with our estimate, 
% which may imply that they are biased by the optically thick dust emission and the line emission from the snow surface.
\revise{which may imply that they predominantly trace the emission from snow surface.}

\textrm{Finally, we briefly discuss the abundance of gas-phase \methanol in the inner region of the V883 Ori disk. A rough estimate of the \methanol abundance at $\lesssim40$\,au can be obtained by comparing the \methanol column density profile and the dust surface density profile as derived by Zhuang et al. (submitted). To convert the dust surface density to the total gas surface density, we assumed a dust-to-gas mass ratio of 0.01, resulting in a total gas column density of $\sim10^{26}$\,cm$^{-2}$ at $\lesssim40$\,au radius (Figure \ref{fig:column_density_profile}). This leads to a \methanol abundance of $\sim10^{-7}$ if we take an averaged \methanol column density of $10^{19}$\,cm$^{-2}$ at $\lesssim40$\,au.} This abundance is similar to or slightly lower than the observed methanol ice abundances in the interstellar medium \citep[e.g.,][]{Boogert2015}, consistent with a scenario where a large fraction of methanol ice has sublimated in the inner disk. However, we note that the abundance estimate here should be treated as a weak lower limit; we assume an optically thin \methanol emission, which can underestimate \methanol column density, and a canonical dust-to-gas mass ratio of 0.01, which can be higher in the inner region where dust grains are accumulated by radial drift, and in turn, result in a lower total gas column density. We need additional observations to better constrain the \methanol abundance, or additional molecular tracers at (sub-)cm wavelength to estimate the molecular abundance ratios, to discuss the chemistry in the inner region of the V883 Ori disk. 
% This would be a scope of upcoming facilities such as next-generation Very Large Array (ngVLA) and Square kilometer Array (SKA).

\section{Summary} \label{sec:summary}
We presented high-resolution ($\approx0\farcs2$) ALMA Band 1 ($\approx7.5$\,mm) observations of \methanol lines in the disk around the outbursting source V883 Ori. We provide direct observational evidence that the \methanol line emission is present in the innermost $\lesssim40$\,au region where previous higher-frequency observations of different molecular lines consistently show an emission cavity. The \methanol radial column density profile, derived by a detailed radially-resolved line profile analysis, indicates a steep increase at $\lesssim40$\,au, suggesting a significant amount of warm \methanol gas in this region. The column density of \methanol reaches $\sim10^{19}$--$10^{20}$ cm$^{-2}$ or even higher at $\sim10$\,au, which is higher than the previous \revise{disk-averaged} estimates using ALMA Bands 6/7 data by a factor of $\gtrsim10$. We therefore suggest that the inner emission cavity seen in the higher-frequency observations is caused by the absorption by the optically thick dust, while the line emission is visible in Band 1 due to a significantly lower dust optical depth. The steep increase in the \methanol line intensity profile could be attributed to the \methanol sublimation radius in the midplane, or the radial temperature structure in the inner region of the disk. 
% We also provide an independent measurement of the \methanol snowline radius of $\sim30$--55\,au, and thus suggest a similar or slightly smaller water snowline radius, 
% supporting the previous inference of 42\,au based on the dust continuum observations but contradicting with the molecular-line-based inference ($\sim75$--120\,au) that are likely biased by the dust absorption.
% The larger snowline radii derived from molecular line observations in higher-frequencies are likely biased by the dust absorption. 
We also roughly estimate the \methanol abundance of $\sim10^{-7}$ with respect to H$_2$ in the $\lesssim40$\,au region, but we need additional observations for a better constraint on the abundance and/or abundance ratios and, in turn, a detailed exploration of chemistry in this region. 

The unique access to sub-cm wavelength provided by ALMA Band 1 allowed us to probe the innermost region of the V883 Ori disk, which is inaccessible due to the large dust optical depth in (sub-)mm wavelengths. A similar observational advantage has been demonstrated in the younger Class~0 source NGC1333~IRAS4A1 \citep{DeSimone2020}, where the line emission of complex organics were detected only at cm wavelengths. While the present 5.5~h observations of \methanol provided essential information to understand the disk physical/chemical structures, significantly deeper observations are required to better constrain the \methanol column density with optically thin lines (or isotopologues) and detect other molecular line tracers in sub-cm wavelengths, which are essential to investigate the chemistry in the inner disk and understand the processes of volatile delivery to planets. Furthermore, sub-cm wavelength observations can probe regions exceeding $\sim300$\,K, where the sublimation of refractory species, including carbonaceous grains and salts, occurs.
These would be primary scopes of the upcoming facilities such as the next-generation Very Large Array (ngVLA) and the Square Kilometer Array (SKA).

%% Please use the acknowledgment and contribution environments. This will 
%% be anonomyized when the "anonymous" style option is used. 
\begin{acknowledgments}
We are grateful to Satoshi Ohashi for fruitful discussion on the effect of the continuum emission on the line emission. 
This paper makes use of the following ALMA data: ADS/JAO.ALMA\#2024.1.01619.S. ALMA is a partnership of ESO (representing its member states), NSF (USA) and NINS (Japan), together with NRC (Canada), NSTC and ASIAA (Taiwan), and KASI (Republic of Korea), in cooperation with the Republic of Chile. The Joint ALMA Observatory is operated by ESO, AUI/NRAO and NAOJ.
Y.Y. is supported by JSPS KAKENHI Grant Number JP23KJ0636 and JP25K23409, and the RIKEN Special Postdoctoral Researcher Program (Fellowships).
S.N. is grateful for support from Grants-in-Aid for the Japan Society for the Promotion of Science (JSPS) Fellows grant No. JP23KJ0329, MEXT/JSPS Grants-in-Aid for Scientific Research (KAKENHI) grant Nos. JP23K13155, JP24K00674, and JP23H05441, and the Start-up Research Grant as one of the University of Tokyo Excellent Young Researchers 2024. N.S. is supported by JSPS KAKENHI Grant Numbers JP20H00182, JP20H05845, and JP26K00737. Data analysis was in part carried out on the Multi-wavelength Data Analysis System operated by the Astronomy Data Center (ADC), National Astronomical Observatory of Japan.
\end{acknowledgments}

\facilities{ALMA}

\software{CASA \citep{CASA}, \texttt{bettermoments} \citep{bettermoments}, \texttt{emcee} \citep{emcee}}


\appendix

\section{Details of Observations and Data Reduction}\label{appendix:observations}
Here we provide the details of ALMA Band 1 observations and Band 6 archival data.

\subsection{Band 1 Observations}
We observed the V883 Ori disk in Band 1 during ALMA Cycle~11 (project code: 2024.1.01619.S, PI: S. Notsu). Observations were performed in seven execution blocks (EBs) with an extended antenna configuration (C-9), resulting in a total on-source integration time of 5.5 hours. Table \ref{tab:observations} summarizes the observational details, including observation dates, number of antennas, on-source integration time, mean precepitable water vapor (PWV), baseline coverage, angular resolution, maximum recoverable scale (MRS), and calibrator information.

The pipeline calibration, including automated self-calibration, was performed using the standard ALMA calibration and imaging pipeline (version 2025.1.0.35) built in the Common Astronomy Software Applications (CASA; \citealt{CASA}) version 6.6.6.17. 
The self-calibration was performed using continuum emission. Only one iteration of phase-only calibration with a solution interval of the EB duration was performed. The solutions were then applied to the spectral line visibilities. We restored the calibrated measurement sets by running the Python script \texttt{scriptForPI.py}, which are used in the subsequent imaging and analysis. 

We first produced a continuum image using the CASA task \texttt{tclean} with Briggs' weighting scheme (\texttt{robust} $=0.5$). We employed the multi-term, multi-frequency synthesis (\texttt{mtmfs}) at a reference frequency of 39.873\,GHz ($\approx7.5$\,mm) with \texttt{nterms} $=2$ and multiscale deconvolver with scales of [$0, 10, 30, 50$] pixels, where the pixel size is 0\farcs01. We used the automasking algorithm implemented in CASA to create masks, and performed deconvolution down to $3\sigma$ level in each iteration (\texttt{nsigma} $=3.0$). 
% Figure \ref{fig:mom0_radial_profile} (left panel) shows the resulting continuum image at 7.5\,mm. 
The resulting beam size and RMS noise level were $0\farcs20\times0\farcs13$ (P.A. $= 61\arcdeg$) and 14\,\textmu Jy\,beam$^{-1}$, respectively. To precisely determine the disk center coordinate, we performed a two-dimensional, single-component Gaussian fit using the CASA task \texttt{imfit} and derived a peak coordinate of $\alpha\mathrm{(ICRS)}= 05^\mathrm{h}38^\mathrm{m}18\fs101$, $\delta\mathrm{(ICRS)} = -07^\mathrm{d}02^\mathrm{m}25\fs99$. This is used as a disk center in the following analysis. 

Several CH$_3$OH lines are covered by the correlator setup, from which we used three CH$_3$OH lines (Table \ref{tab:lines}) that are bright enough for detection. We produced image cubes for these lines using the CASA task \texttt{tclean} with the modified Briggs' weighting scheme (\texttt{briggsbwtaper}, \texttt{robust} $=0.5$) and a velocity channel width of $\approx1$ km s$^{-1}$, covering $\pm50$\,km s$^{-1}$ with respect to the known systemic velocity ($\approx4.3$\,km s$^{-1}$). The channel width is a factor of $\approx2$ larger than the native velocity resolution of \methanol $\nu_t=0$ $J_{K_a,K_c} =$ $7_{0, 7}$--$6_{1, 6}$ line ($\approx0.48$\,km\,s$^{-1}$), while comparable to those of other two lines ($\approx1.1$\,km\,s$^{-1}$). The resulting beam sizes ($\sim0\farcs2$) and noise levels ($\sim10$--20\,K) are listed in Table \ref{tab:lines}. We finally produced the image cubes with a common beam size ($0\farcs21\times0\farcs14$) by smoothing the image cubes using the CASA task \texttt{imsmooth}, which are used in the subsequent analysis.

We note that two additional \methanol lines that have similar intrinsic strengths are covered by other spectral windows, but a significant amount of visibilities in these windows are flagged due to bad data, resulting in a factor of $\approx2$ higher noise level. We thus do not use these two lines in the analysis. 

\subsection{Complimentary Band 6 Archival Data}
We complimentarily used the archival data of \methanol lines in Band 6 (project code: 2021.1.00186.S, PI: J. J. Tobin; \citealt{Tobin2023}). We selected this dataset among the plenty of archival data as its spatial resolution is high enough for the fair comparison at the resolution of the Band 1 data. We make use of the line image cubes available in the Harvard Dataverse Repository \revise{\citep{Tobin_2023_data}}
% \footnote{\url{https://dataverse.harvard.edu/dataset.xhtml?persistentId=doi:10.7910/DVN/MDQJEU}} 
\revise{as described in} \citet{Tobin2023}, where we specifically focused on the unblended \methanol line at 241.852\,GHz ($\nu_\mathrm{t}=0$, $J_{K_a,K_c}=5_{3,2}$--$4_{3,1}$, $E_\mathrm{u}=97.5$\,K). The spatial and velocity resolutions of this data ($\approx0\farcs1$ and $\approx0.05$\,km\,s$^{-1}$) are both higher than the Band 1 data. We aligned the beam size and spatial and velocity grids to those of the Band 1 image cubes using the CASA tasks \texttt{imsmooth} and \texttt{imregrid}, and used them for the subsequent analysis. 

\begin{deluxetable*}{lcccccccc}
\label{tab:observations}
% \tablewidth{\textwidth}
\tablecaption{Observational Details}
\tablehead{\colhead{Date} & \colhead{\# of Ant.} & \colhead{On-source Int.} & \colhead{PWV}  & \colhead{Baseline} & \colhead{Ang. Res.} & \colhead{MRS} & \multicolumn{2}{c}{Calibrators}\\
\cline{8-9}
\colhead{} & \colhead{} & \colhead{(min)} & \colhead{(mm)} & \colhead{(m)} & \colhead{($\arcsec$)} & \colhead{($\arcsec$)} & \colhead{Bandpass/Amplitude} & \colhead{Phase}}
\startdata
2025 Sep. 7  & 43 & 47 & 4.4 & 59.8--15239  & 0.1 & 2.4 & J0423-0120 & J0541-0541 \\
2025 Sep. 13 & 45 & 47 & 1.7 & 59.8--14851  & 0.1 & 2.7 & J0423-0120 & J0541-0541 \\
2025 Sep. 13 & 41 & 47 & 2.0 & 89.4--14851  & 0.1 & 2.7 & J0725-0054 & J0552-0605 \\
2025 Sep. 14 & 43 & 47 & 3.5 & 59.8--14851  & 0.1 & 3.0 & J0423-0120 & J0541-0541 \\
2025 Sep. 14 & 44 & 47 & 3.7 & 59.8--14851  & 0.1 & 3.0 & J0423-0120 & J0541-0541 \\
2025 Sep. 15 & 43 & 47 & 0.9 & 153.1--14851 & 0.1 & 2.4 & J0423-0120 & J0541-0541 \\
2025 Sep. 15 & 43 & 47 & 0.9 & 153.1--14851 & 0.1 & 2.4 & J0725-0054 & J0541-0541 \\
\enddata
\end{deluxetable*}

\section{\revise{Zeroth Moment Maps of Individual \methanol Transitions}}\label{appendix:individual_mom0}
\begin{figure*}
\epsscale{0.85}
\plotone{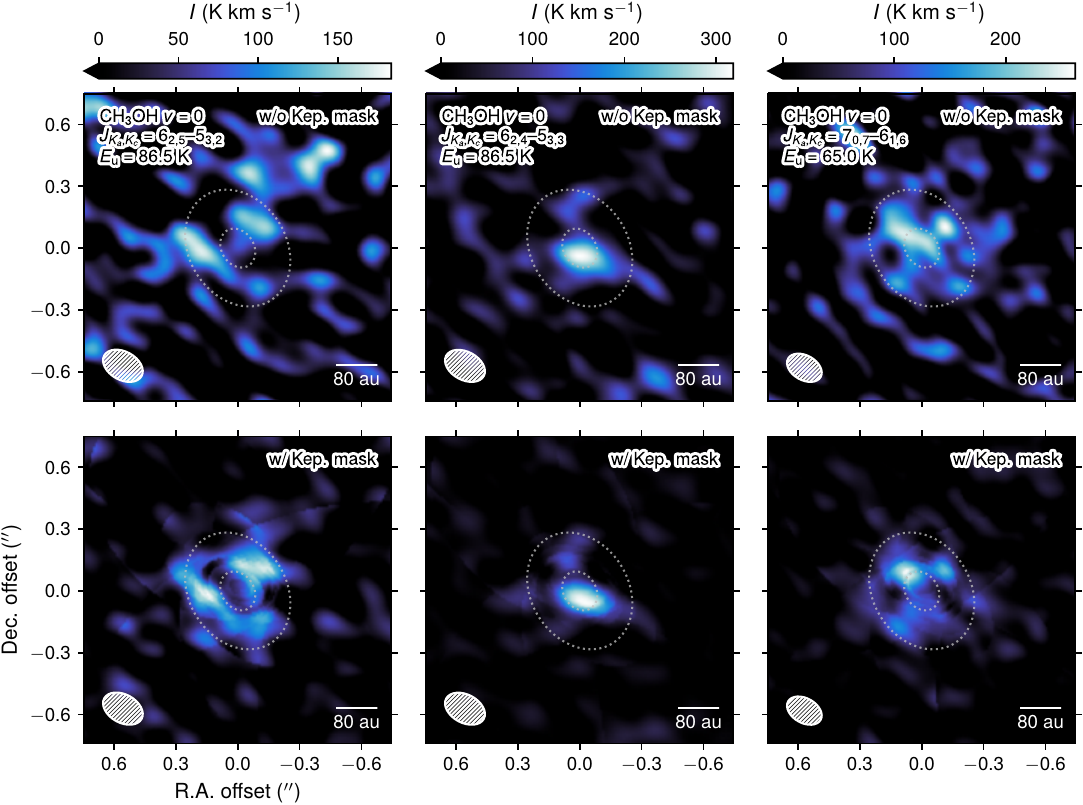}
\caption{\revise{Velocity-integrated intensity maps for individual \methanol transitions with (bottom row) and without (top row) Keplerian mask. The synthesized beam and an 80 au scale bar are located in the lower left and right corners, respectively. The gray dotted ellipses delineate 40 au and 120 au radii.}}
\label{fig:mom0_individual}
\end{figure*}

\revise{Figure \ref{fig:mom0_individual} shows the velocity-integrated intensity (zeroth moment) maps of individual \methanol transitions with and without Keplerian masks, which allows to assess the detections of individual transitions and effect of Keplerian masks that sometime causes artifacts in the maps due to its discrete nature. For the maps without Keplerian mask, we integrate $\pm4$\,km\,s$^{-1}$ with respect to the systemic velocity $v_\mathrm{sys}=4.3$\,km\,s$^{-1}$ \citep[e.g.,][]{Lee2019}. While the maps with Keplerian mask better capture the disk emission, S/Ns appear still low and the maps exhibit patchy structures, which prohibit us to perform detailed analysis. In particular, \methanol $\nu_\mathrm{t}=0$ $J_{K_a,K_c} =$ $6_{2, 4}$--$5_{3, 3}$ line shows a rather stronger, compact emission compared to others, which could either be a noise spike or actual structure. However, due to the limited S/Ns, we here focus on the averaged emission distribution of these three \methanol lines by stacking analysis as described in Section \ref{subsec:distributions}, assuming they trace the same component, and leave the individual line analysis with better sensitivity data for future work.}

\section{Forward Modeling of Radially-resolved Line Profile}\label{appendix:modeling}
Here we describe the methodology to infer the intrinsic radial integrated-intensity profile of the line emission from the observed radial-velocity structures of line intensities (``R-V map''; see Figure \ref{fig:line_profile_fit}). While the direct inference from the observed integrated emission map is limited by a finite spatial resolution and beam convolution, velocity information can allow us to more precisely infer the radial intensity profile assuming a relation between radius $r$ and velocity $v$, which is here Keplerian rotation, i.e., $v_\mathrm{Kep}(r) = \sqrt{GM_\star/r}\sin i$. To do so, we employed a forward modeling approach that fully takes the beam smearing effect into account. 
% Practically, we compute a model line cube of a simple Keplerian-rotating disk for a given radial integrated-intensity profile $I(r)$ (see Equations \ref{eq:model_cube} and \ref{eq:los_velocity}), apply a beam convolution, and then extract the model R-V map which is fitted to the observed R-V map. 
We first represent the radial integrated-intensity profile $I(r)$ as a discretized function using a set of basis functions:
\begin{equation}
    I(r) = \sum_{i=1}^{n_r}I_i\varphi_i(r),
\end{equation}
where $\varphi_i(r)$ is a basis function 
\begin{equation}
    \varphi_i(r) = 
    \begin{cases}
        1 & (r_{i-1}\leq r < r_{i}), \\
        0 & \text{(otherwise)}, 
    \end{cases}
\end{equation}
and $I_i$ is the integrated intensity within the corresponding radial region ($r_{i-1}\leq r < r_{i}$) with $n_r$ being the number of radial grid $r_i$. The unknown free parameters we are going to estimate are thus given by the coefficient vector
\begin{equation}
    \boldsymbol{I} = (I_1, I_2, \ldots, I_{n_r})^\mathrm{T}.
\end{equation}

The radial grid spacing $\Delta r$ is chosen to avoid oversampling that can result in an ill-posed problem while fully utilizing the available spatial and spectral information. Since the radial spacing corresponding to a given velocity spacing $\Delta v$ is approximately given by $|\mathrm{d}r/\mathrm{d}v|\Delta v \approx 2r\Delta v/v_\mathrm{Kep}$, we specifically adopt 
\begin{equation}
    \Delta r = \min\left\{\alpha\frac{2r\Delta v}{v_\mathrm{Kep}}, \frac{\theta_\mathrm{beam}}{2}\right\},
\end{equation}
where $\theta_\mathrm{beam}$ is the full width of half maximum of the beam major axis ($\approx0\farcs21$, or $\approx80$\,au), and $\alpha$ is an arbitrary parameter that controls the sampling relative to the velocity resolution and depends on the S/Ns. We adopt $\Delta v = 1$\,km\,s$^{-1}$ based on the data channel width, and $\alpha=2$ based on empirical tests. The innermost and outermost radii are chosen to be $r_0 = 10$\,au and $r_{n_r} = 350$\,au based on the radial and velocity distribution of the emission (Figures \ref{fig:mom0_radial_profile}).
% , resulting in a total of eleven radial samplings (and thus free parameters).

We numerically compute the model line cube with a grid of deprojected disk coordinate $(r,\phi)$ following Equation (\ref{eq:model_cube}), apply Gaussian convolutions to consider the spatial and spectral resolutions, and extract a model R-V map in the same manner as the data, which is then fitted to the observed R-V map. However, computing the full model, particularly the Gaussian beam convolution, is highly computationally expensive for the fitting process. We instead pre-compute the beam response to the unit intensities $\boldsymbol{I} = (1, 1, \ldots, 1)^\mathrm{T}$ at different data points of the R-V map to construct an $n_\mathrm{data}\times n_r$ response matrix $\boldsymbol{W}$, where $n_\mathrm{data}$ is the number of data points of the R-V map. Using this response matrix, we construct a likelihood function,
\begin{align}
    &\mathcal{L}(\boldsymbol{D}\,|\,\boldsymbol{I}) \propto \exp\left(-\frac{1}{2}(\boldsymbol{WI} - \boldsymbol{D})^\mathrm{T}\boldsymbol{C}^{-1}(\boldsymbol{WI} - \boldsymbol{D})\right),
\end{align}
where $\boldsymbol{D}$ is the vector representing the observed R-V map (flattened), and $\boldsymbol{C}$ is the diagonal covariance matrix,
\begin{equation}
\boldsymbol{C}
=
\begin{pmatrix}
\sigma_1^2 & & &  \\
& \sigma_2^2 & &  \\
& & \ddots & \\
& & & \sigma_{n_\mathrm{data}}^2
\end{pmatrix},
\end{equation}
with $\sigma_i\,(i=1,2,\ldots,n_\mathrm{data})$ being the statistical uncertainties of the observed R-V map. 

% Further, we consider the flux calibration uncertainty, which is expected to be $\sim5\%$ at Band 1, in the following way. We treat this systematic uncertainty as a fully  spatially correlated noise on the R-V map, and introduce a constant factor $C$ that represent the flux calibration uncertainty 

We compute the posterior distributions of $\boldsymbol{I}$ using the affine-invariant MCMC ensemble sampler in the Python package \texttt{emcee} \citep{emcee}. We use 100 walkers, each taking 5,000 steps for burn-in and additional 5,000 steps to sample the posterior distributions. To determine the priors and the initial positions of walkers, we solve the minimization problem, $\arg\min_{\boldsymbol{I}} \left\| \boldsymbol{WI} - \boldsymbol{D} \right\|_2^2$ for $\boldsymbol{I}\geq0$, using the non-negative least square solver implemented in \texttt{scipy.optimize.nnls}. We use the solutions as the starting points of walkers, and impose an uninformative (uniform) prior of $\mathcal{U}(0, 10^4)$ common for all $I_i$ in K\,km\,s$^{-1}$. 

We performed this sort of analysis for the stacked \methanol line in Band 1 and a single \methanol line in Band 6, and show the results in Figure \ref{fig:line_profile_fit}. For the sake of the visualization of the radial intensity profiles, we plot the Piecewise Cubic Hermite Interpolating Polynomial (PCHIP) of the constrained data points, in addition to the values of $\boldsymbol{I}$, in Figure \ref{fig:line_profile_fit} (right).

\section{\revise{Comparison of the radial intensity profiles derived from line profile fit and image deprojection}}\label{appendix:profile_comparison}

\begin{figure*}
\epsscale{0.9}
\plotone{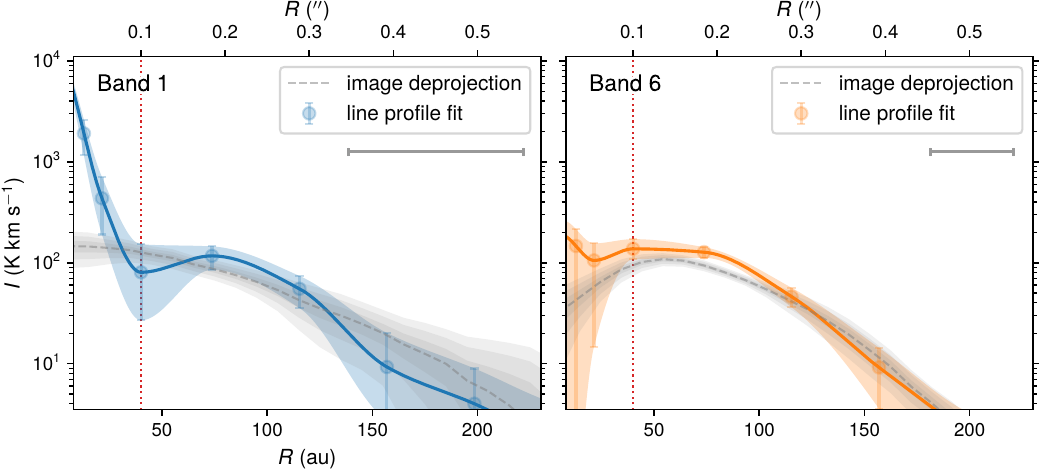}
\caption{\revise{Comparison of the Band 1 (left) and Band 6 (right) \methanol radial intensity profiles derived from the line profile fit (color) and from the image deprojection (gray). The color-shaded regions indicate the 68\% highest density interval of the MCMC posterior distributions, while the gray-shaded regions exhibit [1, 2, 3]$\sigma$ uncertainties associated with the image-derived profiles. The beam major axes of the images used for deprojection are indicated by the horizontal bars in the top right corners. The vertical red dotted lines mark 40 au radius.}}
\label{fig:radial_profile_comparison}
\end{figure*}

\revise{Figure \ref{fig:radial_profile_comparison} shows the comparisons of the radial intensity profiles derived from the line profile fit (Section \ref{subsec:line_profile_fit}) and image deprojection (Section \ref{subsec:distributions}). 
% To better visualize the overall morphologies of the profiles, we show these comparisons in semi-log plot. 
Overall, the line profile fit reproduces the fine radial structures while the image deprojection is predominantly limited by the finite spatial resolutions. The intensity profile of the Band 6 \methanol line from the line profile fit is slightly offset compared to the profile directly from the deprojection of the high-resolution ($\approx0\farcs1$) image, which is consistent each other considering the beam dilution effect, validating the line profile fit method. \revise{We note that the reconstructed radial intensity profile of a \methanol transition in Band 3 also show a similar distribution as the Band 6 profiles \citep{Yamato2024_V883Ori}.}
}

%% For this sample we use BibTeX plus aasjournalv7.bst to generate the

\bibliography{reference}{}

@ARTICLE{bettermoments,
       author = {{Teague}, Richard and {Foreman-Mackey}, Daniel},
        title = "{A Robust Method to Measure Centroids of Spectral Lines}",
      journal = {Research Notes of the American Astronomical Society},
         year = 2018,
        month = Sep,
       volume = {2},
          eid = {173},
        pages = {173},
          doi = {10.3847/2515-5172/aae265},
       adsurl = {https://ui.adsabs.harvard.edu/abs/2018RNAAS...2c.173T}
}

@ARTICLE{CASA,
       author = {{CASA Team} and {Bean}, Ben and {Bhatnagar}, Sanjay and {Castro}, Sandra and {Donovan Meyer}, Jennifer and {Emonts}, Bjorn and {Garcia}, Enrique and {Garwood}, Robert and {Golap}, Kumar and {Gonzalez Villalba}, Justo and {Harris}, Pamela and {Hayashi}, Yohei and {Hoskins}, Josh and {Hsieh}, Mingyu and {Jagannathan}, Preshanth and {Kawasaki}, Wataru and {Keimpema}, Aard and {Kettenis}, Mark and {Lopez}, Jorge and {Marvil}, Joshua and {Masters}, Joseph and {McNichols}, Andrew and {Mehringer}, David and {Miel}, Renaud and {Moellenbrock}, George and {Montesino}, Federico and {Nakazato}, Takeshi and {Ott}, Juergen and {Petry}, Dirk and {Pokorny}, Martin and {Raba}, Ryan and {Rau}, Urvashi and {Schiebel}, Darrell and {Schweighart}, Neal and {Sekhar}, Srikrishna and {Shimada}, Kazuhiko and {Small}, Des and {Steeb}, Jan-Willem and {Sugimoto}, Kanako and {Suoranta}, Ville and {Tsutsumi}, Takahiro and {van Bemmel}, Ilse M. and {Verkouter}, Marjolein and {Wells}, Akeem and {Xiong}, Wei and {Szomoru}, Arpad and {Griffith}, Morgan and {Glendenning}, Brian and {Kern}, Jeff},
        title = "{CASA, the Common Astronomy Software Applications for Radio Astronomy}",
      journal = {\pasp},
         year = 2022,
        month = nov,
       volume = {134},
       number = {1041},
          eid = {114501},
        pages = {114501},
          doi = {10.1088/1538-3873/ac9642},
archivePrefix = {arXiv},
       eprint = {2210.02276},
 primaryClass = {astro-ph.IM},
       adsurl = {https://ui.adsabs.harvard.edu/abs/2022PASP..134k4501C}
}

@ARTICLE{CDMS1,
       author = {{M{\"u}ller}, H.~S.~P. and {Thorwirth}, S. and {Roth}, D.~A. and {Winnewisser}, G.},
        title = "{The Cologne Database for Molecular Spectroscopy, CDMS}",
      journal = {\aap},
         year = 2001,
        month = apr,
       volume = {370},
        pages = {L49-L52},
          doi = {10.1051/0004-6361:20010367},
       adsurl = {https://ui.adsabs.harvard.edu/abs/2001A&A...370L..49M}
}

@ARTICLE{CDMS2,
       author = {{M{\"u}ller}, Holger S.~P. and {Schl{\"o}der}, Frank and {Stutzki}, J{\"u}rgen and {Winnewisser}, Gisbert},
        title = "{The Cologne Database for Molecular Spectroscopy, CDMS: a useful tool for astronomers and spectroscopists}",
      journal = {Journal of Molecular Structure},
         year = 2005,
        month = may,
       volume = {742},
       number = {1-3},
        pages = {215-227},
          doi = {10.1016/j.molstruc.2005.01.027},
       adsurl = {https://ui.adsabs.harvard.edu/abs/2005JMoSt.742..215M}
}

@ARTICLE{CDMS3,
       author = {{Endres}, Christian P. and {Schlemmer}, Stephan and {Schilke}, Peter and {Stutzki}, J{\"u}rgen and {M{\"u}ller}, Holger S.~P.},
        title = "{The Cologne Database for Molecular Spectroscopy, CDMS, in the Virtual Atomic and Molecular Data Centre, VAMDC}",
      journal = {Journal of Molecular Spectroscopy},
         year = 2016,
        month = sep,
       volume = {327},
        pages = {95-104},
          doi = {10.1016/j.jms.2016.03.005},
       adsurl = {https://ui.adsabs.harvard.edu/abs/2016JMoSp.327...95E}
}

@ARTICLE{Lees1968,
       author = {{Lees}, R.~M. and {Baker}, J.~G.},
        title = "{Torsion-Vibration-Rotation Interactions in Methanol. I. Millimeter Wave Spectrum}",
      journal = {\jcp},
         year = 1968,
        month = jun,
       volume = {48},
       number = {12},
        pages = {5299-5318},
          doi = {10.1063/1.1668221},
       adsurl = {https://ui.adsabs.harvard.edu/abs/1968JChPh..48.5299L}
}

@ARTICLE{Cieza2016,
       author = {{Cieza}, Lucas A. and {Casassus}, Simon and {Tobin}, John and {Bos}, Steven P. and {Williams}, Jonathan P. and {Perez}, Sebastian and {Zhu}, Zhaohuan and {Caceres}, Claudio and {Canovas}, Hector and {Dunham}, Michael M. and {Hales}, Antonio and {Prieto}, Jose L. and {Principe}, David A. and {Schreiber}, Matthias R. and {Ruiz-Rodriguez}, Dary and {Zurlo}, Alice},
        title = "{Imaging the water snow-line during a protostellar outburst}",
      journal = {\nat},
         year = 2016,
        month = jul,
       volume = {535},
       number = {7611},
        pages = {258-261},
          doi = {10.1038/nature18612},
archivePrefix = {arXiv},
       eprint = {1607.03757},
 primaryClass = {astro-ph.SR},
       adsurl = {https://ui.adsabs.harvard.edu/abs/2016Natur.535..258C}
}

@ARTICLE{Tobin2023,
       author = {{Tobin}, John J. and {van't Hoff}, Merel L.~R. and {Leemker}, Margot and {van Dishoeck}, Ewine F. and {Paneque-Carre{\~n}o}, Teresa and {Furuya}, Kenji and {Harsono}, Daniel and {Persson}, Magnus V. and {Cleeves}, L. Ilsedore and {Sheehan}, Patrick D. and {Cieza}, Lucas},
        title = "{Deuterium-enriched water ties planet-forming disks to comets and protostars}",
      journal = {\nat},
         year = 2023,
        month = mar,
       volume = {615},
       number = {7951},
        pages = {227-230},
          doi = {10.1038/s41586-022-05676-z},
       adsurl = {https://ui.adsabs.harvard.edu/abs/2023Natur.615..227T}
}

@ARTICLE{vantHoff2018,
       author = {{van 't Hoff}, Merel L.~R. and {Tobin}, John J. and {Trapman}, Leon and {Harsono}, Daniel and {Sheehan}, Patrick D. and {Fischer}, William J. and {Megeath}, S. Thomas and {van Dishoeck}, Ewine F.},
        title = "{Methanol and its Relation to the Water Snowline in the Disk around the Young Outbursting Star V883 Ori}",
      journal = {\apjl},
         year = 2018,
        month = sep,
       volume = {864},
       number = {1},
          eid = {L23},
        pages = {L23},
          doi = {10.3847/2041-8213/aadb8a},
archivePrefix = {arXiv},
       eprint = {1808.08258},
 primaryClass = {astro-ph.SR},
       adsurl = {https://ui.adsabs.harvard.edu/abs/2018ApJ...864L..23V}
}

@ARTICLE{Lee2019,
       author = {{Lee}, Jeong-Eun and {Lee}, Seokho and {Baek}, Giseon and {Aikawa}, Yuri and {Cieza}, Lucas and {Yoon}, Sung-Yong and {Herczeg}, Gregory and {Johnstone}, Doug and {Casassus}, Simon},
        title = "{The ice composition in the disk around V883 Ori revealed by its stellar outburst}",
      journal = {Nature Astronomy},
         year = 2019,
        month = feb,
       volume = {3},
        pages = {314-319},
          doi = {10.1038/s41550-018-0680-0},
archivePrefix = {arXiv},
       eprint = {1809.00353},
 primaryClass = {astro-ph.SR},
       adsurl = {https://ui.adsabs.harvard.edu/abs/2019NatAs...3..314L}
}

@ARTICLE{emcee,
       author = {{Foreman-Mackey}, Daniel and {Hogg}, David W. and {Lang}, Dustin and {Goodman}, Jonathan},
        title = "{emcee: The MCMC Hammer}",
      journal = {\pasp},
         year = 2013,
        month = mar,
       volume = {125},
       number = {925},
        pages = {306},
          doi = {10.1086/670067},
archivePrefix = {arXiv},
       eprint = {1202.3665},
 primaryClass = {astro-ph.IM},
       adsurl = {https://ui.adsabs.harvard.edu/abs/2013PASP..125..306F}
}

@ARTICLE{Notsu2021,
       author = {{Notsu}, Shota and {van Dishoeck}, Ewine F. and {Walsh}, Catherine and {Bosman}, Arthur D. and {Nomura}, Hideko},
        title = "{X-ray-induced chemistry of water and related molecules in low-mass protostellar envelopes}",
      journal = {\aap},
         year = 2021,
        month = jun,
       volume = {650},
          eid = {A180},
        pages = {A180},
          doi = {10.1051/0004-6361/202140667},
archivePrefix = {arXiv},
       eprint = {2104.06878},
 primaryClass = {astro-ph.GA},
       adsurl = {https://ui.adsabs.harvard.edu/abs/2021A&A...650A.180N}
}

@ARTICLE{Bosman2021,
       author = {{Bosman}, Arthur D. and {Bergin}, Edwin A. and {Loomis}, Ryan A. and {Andrews}, Sean M. and {van't Hoff}, Merel L.~R. and {Teague}, Richard and {{\"O}berg}, Karin I. and {Guzm{\'a}n}, Viviana V. and {Walsh}, Catherine and {Aikawa}, Yuri and {Alarc{\'o}n}, Felipe and {Bae}, Jaehan and {Bergner}, Jennifer B. and {Booth}, Alice S. and {Cataldi}, Gianni and {Cleeves}, L. Ilsedore and {Czekala}, Ian and {Huang}, Jane and {Ilee}, John D. and {Law}, Charles J. and {Le Gal}, Romane and {Liu}, Yao and {Long}, Feng and {M{\'e}nard}, Fran{\c{c}}ois and {Nomura}, Hideko and {P{\'e}rez}, Laura M. and {Qi}, Chunhua and {Schwarz}, Kamber R. and {Sierra}, Anibal and {Tsukagoshi}, Takashi and {Yamato}, Yoshihide and {Wilner}, David J. and {Zhang}, Ke},
        title = "{Molecules with ALMA at Planet-forming Scales (MAPS). XV. Tracing Protoplanetary Disk Structure within 20 au}",
      journal = {\apjs},
         year = 2021,
        month = nov,
       volume = {257},
       number = {1},
          eid = {15},
        pages = {15},
          doi = {10.3847/1538-4365/ac1433},
archivePrefix = {arXiv},
       eprint = {2109.06223},
 primaryClass = {astro-ph.EP},
       adsurl = {https://ui.adsabs.harvard.edu/abs/2021ApJS..257...15B}
}

@ARTICLE{Boogert2015,
       author = {{Boogert}, A.~C. Adwin and {Gerakines}, Perry A. and {Whittet}, Douglas C.~B.},
        title = "{Observations of the icy universe.}",
      journal = {\araa},
         year = 2015,
        month = aug,
       volume = {53},
        pages = {541-581},
          doi = {10.1146/annurev-astro-082214-122348},
archivePrefix = {arXiv},
       eprint = {1501.05317},
 primaryClass = {astro-ph.GA},
       adsurl = {https://ui.adsabs.harvard.edu/abs/2015ARA&A..53..541B}
}

@ARTICLE{Eistrup2016,
       author = {{Eistrup}, Christian and {Walsh}, Catherine and {van Dishoeck}, Ewine F.},
        title = "{Setting the volatile composition of (exo)planet-building material. Does chemical evolution in disk midplanes matter?}",
      journal = {\aap},
         year = 2016,
        month = nov,
       volume = {595},
          eid = {A83},
        pages = {A83},
          doi = {10.1051/0004-6361/201628509},
archivePrefix = {arXiv},
       eprint = {1607.06710},
 primaryClass = {astro-ph.EP},
       adsurl = {https://ui.adsabs.harvard.edu/abs/2016A&A...595A..83E}
}

@ARTICLE{Strom1993,
       author = {{Strom}, Karen M. and {Strom}, Stephen E.},
        title = "{The Discovery of Two FU Orionis Objects in L1641}",
      journal = {\apjl},
         year = 1993,
        month = aug,
       volume = {412},
        pages = {L63},
          doi = {10.1086/186941},
archivePrefix = {arXiv},
       eprint = {astro-ph/9305012},
 primaryClass = {astro-ph},
       adsurl = {https://ui.adsabs.harvard.edu/abs/1993ApJ...412L..63S}
}

@ARTICLE{Furlan2016,
       author = {{Furlan}, E. and {Fischer}, W.~J. and {Ali}, B. and {Stutz}, A.~M. and {Stanke}, T. and {Tobin}, J.~J. and {Megeath}, S.~T. and {Osorio}, M. and {Hartmann}, L. and {Calvet}, N. and {Poteet}, C.~A. and {Booker}, J. and {Manoj}, P. and {Watson}, D.~M. and {Allen}, L.},
        title = "{The Herschel Orion Protostar Survey: Spectral Energy Distributions and Fits Using a Grid of Protostellar Models}",
      journal = {\apjs},
         year = 2016,
        month = may,
       volume = {224},
       number = {1},
          eid = {5},
        pages = {5},
          doi = {10.3847/0067-0049/224/1/5},
archivePrefix = {arXiv},
       eprint = {1602.07314},
 primaryClass = {astro-ph.SR},
       adsurl = {https://ui.adsabs.harvard.edu/abs/2016ApJS..224....5F}
}

@ARTICLE{Lee2023,
       author = {{Lee}, Jeong-Eun and {Baek}, Giseon and {Lee}, Seokho and {Jeong}, Jae-Hong and {Kim}, Chul-Hwan and {Aikawa}, Yuri and {Herczeg}, Gregory J. and {Johnstone}, Doug and {Tobin}, John J.},
        title = "{Complex Organic Molecules in a Very Young Hot Corino, HOPS 373SW}",
      journal = {arXiv e-prints},
         year = 2023,
        month = jun,
          eid = {arXiv:2306.16959},
        pages = {arXiv:2306.16959},
          doi = {10.48550/arXiv.2306.16959},
archivePrefix = {arXiv},
       eprint = {2306.16959},
 primaryClass = {astro-ph.SR},
       adsurl = {https://ui.adsabs.harvard.edu/abs/2023arXiv230616959L}
}

@ARTICLE{Yamato2024_V883Ori,
       author = {{Yamato}, Yoshihide and {Notsu}, Shota and {Aikawa}, Yuri and {Okoda}, Yuki and {Nomura}, Hideko and {Sakai}, Nami},
        title = "{Chemistry of Complex Organic Molecules in the V883 Ori Disk Revealed by ALMA Band 3 Observations}",
      journal = {\aj},
         year = 2024,
        month = feb,
       volume = {167},
       number = {2},
          eid = {66},
        pages = {66},
          doi = {10.3847/1538-3881/ad11d9},
archivePrefix = {arXiv},
       eprint = {2312.01300},
 primaryClass = {astro-ph.EP},
       adsurl = {https://ui.adsabs.harvard.edu/abs/2024AJ....167...66Y}
}

@ARTICLE{Jeong2025,
       author = {{Jeong}, Jae-Hong and {Lee}, Jeong-Eun and {Lee}, Seonjae and {Baek}, Giseon and {Kang}, Ji-Hyun and {Lee}, Seokho and {Kim}, Chul-Hwan and {Yun}, Hyeong-Sik and {Aikawa}, Yuri and {Herczeg}, Gregory J. and {Johnstone}, Doug and {Cieza}, Lucas},
        title = "{ALMA Spectral Survey of an Eruptive Young Star, V883 Ori (ASSAY). II. Freshly Sublimated Complex Organic Molecules in the Keplerian Disk}",
      journal = {\apjs},
         year = 2025,
        month = feb,
       volume = {276},
       number = {2},
          eid = {49},
        pages = {49},
          doi = {10.3847/1538-4365/ad9450},
archivePrefix = {arXiv},
       eprint = {2411.03826},
 primaryClass = {astro-ph.SR},
       adsurl = {https://ui.adsabs.harvard.edu/abs/2025ApJS..276...49J}
}

@ARTICLE{Leemker2021,
       author = {{Leemker}, M. and {van't Hoff}, M.~L.~R. and {Trapman}, L. and {van Gelder}, M.~L. and {Hogerheijde}, M.~R. and {Ru{\'\i}z-Rodr{\'\i}guez}, D. and {van Dishoeck}, E.~F.},
        title = "{Chemically tracing the water snowline in protoplanetary disks with HCO$^{+}$}",
      journal = {\aap},
         year = 2021,
        month = feb,
       volume = {646},
          eid = {A3},
        pages = {A3},
          doi = {10.1051/0004-6361/202039387},
archivePrefix = {arXiv},
       eprint = {2011.12319},
 primaryClass = {astro-ph.EP},
       adsurl = {https://ui.adsabs.harvard.edu/abs/2021A&A...646A...3L}
}

@ARTICLE{Houge2024,
       author = {{Houge}, Adrien and {Mac{\'\i}as}, Enrique and {Krijt}, Sebastiaan},
        title = "{Surviving the heat: multiwavelength analysis of V883 Ori reveals that dust aggregates survive the sublimation of their ice mantles}",
      journal = {\mnras},
         year = 2024,
        month = feb,
       volume = {527},
       number = {4},
        pages = {9668-9682},
          doi = {10.1093/mnras/stad3758},
archivePrefix = {arXiv},
       eprint = {2312.01856},
 primaryClass = {astro-ph.EP},
       adsurl = {https://ui.adsabs.harvard.edu/abs/2024MNRAS.527.9668H}
}

@phdthesis{Leemker2024_thesis,
  author       = {Leemker, Margot},
  title        = {Freezing conditions in warm disks: snowlines and their effect on the chemical structure of planet-forming disks},
  school       = {Leiden UNiversity},
  year         = {2024}
}

@ARTICLE{Minissale2022,
       author = {{Minissale}, Marco and {Aikawa}, Yuri and {Bergin}, Edwin and {Bertin}, Mathieu and {Brown}, Wendy A. and {Cazaux}, Stephanie and {Charnley}, Steven B. and {Coutens}, Audrey and {Cuppen}, Herma M. and {Guzman}, Victoria and {Linnartz}, Harold and {McCoustra}, Martin R.~S. and {Rimola}, Albert and {Schrauwen}, Johanna G.~M. and {Toubin}, Celine and {Ugliengo}, Piero and {Watanabe}, Naoki and {Wakelam}, Valentine and {Dulieu}, Francois},
        title = "{Thermal Desorption of Interstellar Ices: A Review on the Controlling Parameters and Their Implications from Snowlines to Chemical Complexity}",
      journal = {ACS Earth and Space Chemistry},
         year = 2022,
        month = mar,
       volume = {6},
       number = {3},
        pages = {597-630},
          doi = {10.1021/acsearthspacechem.1c00357},
archivePrefix = {arXiv},
       eprint = {2201.07512},
 primaryClass = {astro-ph.GA},
       adsurl = {https://ui.adsabs.harvard.edu/abs/2022ESC.....6..597M}
}

@ARTICLE{Oberg2011,
       author = {{{\"O}berg}, Karin I. and {Murray-Clay}, Ruth and {Bergin}, Edwin A.},
        title = "{The Effects of Snowlines on C/O in Planetary Atmospheres}",
      journal = {\apjl},
         year = 2011,
        month = dec,
       volume = {743},
       number = {1},
          eid = {L16},
        pages = {L16},
          doi = {10.1088/2041-8205/743/1/L16},
archivePrefix = {arXiv},
       eprint = {1110.5567},
 primaryClass = {astro-ph.GA},
       adsurl = {https://ui.adsabs.harvard.edu/abs/2011ApJ...743L..16O}
}

@ARTICLE{Lee2024,
       author = {{Lee}, Jeong-Eun and {Kim}, Chul-Hwan and {Lee}, Seokho and {Lee}, Seonjae and {Baek}, Giseon and {Yun}, Hyeong-Sik and {Aikawa}, Yuri and {Johnstone}, Doug and {Herczeg}, Gregory J. and {Cieza}, Lucas},
        title = "{ALMA Spectral Survey of an Eruptive Young Star, V883 Ori (ASSAY). I. What Triggered the Current Episode of Eruption?}",
      journal = {\apj},
         year = 2024,
        month = may,
       volume = {966},
       number = {1},
          eid = {119},
        pages = {119},
          doi = {10.3847/1538-4357/ad3106},
archivePrefix = {arXiv},
       eprint = {2403.03436},
 primaryClass = {astro-ph.SR},
       adsurl = {https://ui.adsabs.harvard.edu/abs/2024ApJ...966..119L}
}

@ARTICLE{Fadul2024a,
       author = {{Fadul}, Abubakar M.~A. and {Schwarz}, Kamber R. and {van'T Hoff}, Merel L.~R. and {Huang}, Jane and {Bergner}, Jennifer B. and {Suhasaria}, Tushar and {Calahan}, Jenny K.},
        title = "{A Deep Search for Complex Organic Molecules toward the Protoplanetary Disk of V883 Ori}",
      journal = {\aj},
         year = 2025,
        month = jun,
       volume = {169},
       number = {6},
          eid = {307},
        pages = {307},
          doi = {10.3847/1538-3881/adc998},
archivePrefix = {arXiv},
       eprint = {2504.06005},
 primaryClass = {astro-ph.SR},
       adsurl = {https://ui.adsabs.harvard.edu/abs/2025AJ....169..307F}
}

@ARTICLE{Fadul2024b,
       author = {{Fadul}, Abubakar M.~A. and {Schwarz}, Kamber R. and {Suhasaria}, Tushar and {Calahan}, Jenny K. and {Huang}, Jane and {van't Hoff}, Merel L.~R.},
        title = "{A Deep Search for Ethylene Glycol and Glycolonitrile in the V883 Ori Protoplanetary Disk}",
      journal = {\apjl},
         year = 2025,
        month = aug,
       volume = {988},
       number = {2},
          eid = {L44},
        pages = {L44},
          doi = {10.3847/2041-8213/adec6e},
archivePrefix = {arXiv},
       eprint = {2507.14905},
 primaryClass = {astro-ph.SR},
       adsurl = {https://ui.adsabs.harvard.edu/abs/2025ApJ...988L..44F}
}

@ARTICLE{Leemker2025,
       author = {{Leemker}, Margot and {Tobin}, John J. and {Facchini}, Stefano and {Curone}, Pietro and {Booth}, Alice S. and {Furuya}, Kenji and {van't Hoff}, Merel L.~R.},
        title = "{Pristine ices in a planet-forming disk revealed by heavy water}",
      journal = {Nature Astronomy},
         year = 2025,
        month = oct,
       volume = {9},
        pages = {1486-1494},
          doi = {10.1038/s41550-025-02663-y},
archivePrefix = {arXiv},
       eprint = {2510.19919},
 primaryClass = {astro-ph.EP},
       adsurl = {https://ui.adsabs.harvard.edu/abs/2025NatAs...9.1486L}
}

@ARTICLE{Zeng2025,
       author = {{Zeng}, Shaoshan and {Jeong}, Jae-Hong and {Oyama}, Takahiro and {Lee}, Jeong-Eun and {Yang}, Yao-Lun and {Sakai}, Nami},
        title = "{Determining the Methanol Deuteration in the Disk Around V883 Orionis with Laboratory Measured Spectroscopy}",
      journal = {\aj},
         year = 2025,
        month = jul,
       volume = {170},
       number = {1},
          eid = {33},
        pages = {33},
          doi = {10.3847/1538-3881/add733},
archivePrefix = {arXiv},
       eprint = {2506.07794},
 primaryClass = {astro-ph.SR},
       adsurl = {https://ui.adsabs.harvard.edu/abs/2025AJ....170...33Z}
}

@ARTICLE{Nakasone2026,
       author = {{Nakasone}, Hiroto and {Notsu}, Shota and {Yoshida}, Tomohiro C. and {Nomura}, Hideko and {Tsukagoshi}, Takashi and {Hirota}, Tomoya and {Honda}, Mitsuhiko and {Akiyama}, Eiji and {Booth}, Alice S. and {Lee}, Jeong-Eun and {Lee}, Seokho},
        title = "{ALMA Band 7 Observations of Water Lines in the Protoplanetary Disk of V883 Ori}",
      journal = {\apj},
         year = 2026,
        month = feb,
       volume = {998},
       number = {1},
          eid = {53},
        pages = {53},
          doi = {10.3847/1538-4357/ae2c82},
archivePrefix = {arXiv},
       eprint = {2512.15108},
 primaryClass = {astro-ph.EP},
       adsurl = {https://ui.adsabs.harvard.edu/abs/2026ApJ...998...53N}
}

@ARTICLE{Birnstiel2010,
       author = {{Birnstiel}, T. and {Dullemond}, C.~P. and {Brauer}, F.},
        title = "{Gas- and dust evolution in protoplanetary disks}",
      journal = {\aap},
         year = 2010,
        month = apr,
       volume = {513},
          eid = {A79},
        pages = {A79},
          doi = {10.1051/0004-6361/200913731},
archivePrefix = {arXiv},
       eprint = {1002.0335},
 primaryClass = {astro-ph.EP},
       adsurl = {https://ui.adsabs.harvard.edu/abs/2010A&A...513A..79B}
}

@ARTICLE{Banzatti2015,
       author = {{Banzatti}, A. and {Pinilla}, P. and {Ricci}, L. and {Pontoppidan}, K.~M. and {Birnstiel}, T. and {Ciesla}, F.},
        title = "{Direct Imaging of the Water Snow Line at the Time of Planet Formation using Two ALMA Continuum Bands}",
      journal = {\apjl},
         year = 2015,
        month = dec,
       volume = {815},
       number = {1},
          eid = {L15},
        pages = {L15},
          doi = {10.1088/2041-8205/815/1/L15},
archivePrefix = {arXiv},
       eprint = {1511.06762},
 primaryClass = {astro-ph.EP},
       adsurl = {https://ui.adsabs.harvard.edu/abs/2015ApJ...815L..15B}
}

@ARTICLE{Pinilla2016,
       author = {{Pinilla}, P. and {Klarmann}, L. and {Birnstiel}, T. and {Benisty}, M. and {Dominik}, C. and {Dullemond}, C.~P.},
        title = "{A tunnel and a traffic jam: How transition disks maintain a detectable warm dust component despite the presence of a large planet-carved gap}",
      journal = {\aap},
         year = 2016,
        month = jan,
       volume = {585},
          eid = {A35},
        pages = {A35},
          doi = {10.1051/0004-6361/201527131},
archivePrefix = {arXiv},
       eprint = {1511.04105},
 primaryClass = {astro-ph.EP},
       adsurl = {https://ui.adsabs.harvard.edu/abs/2016A&A...585A..35P}
}

@ARTICLE{Schoonenberg2017,
       author = {{Schoonenberg}, Djoeke and {Okuzumi}, Satoshi and {Ormel}, Chris W.},
        title = "{What pebbles are made of: Interpretation of the V883 Ori disk}",
      journal = {\aap},
         year = 2017,
        month = sep,
       volume = {605},
          eid = {L2},
        pages = {L2},
          doi = {10.1051/0004-6361/201731202},
archivePrefix = {arXiv},
       eprint = {1708.03328},
 primaryClass = {astro-ph.EP},
       adsurl = {https://ui.adsabs.harvard.edu/abs/2017A&A...605L...2S}
}

@ARTICLE{Oka2011,
       author = {{Oka}, Akinori and {Nakamoto}, Taishi and {Ida}, Shigeru},
        title = "{Evolution of Snow Line in Optically Thick Protoplanetary Disks: Effects of Water Ice Opacity and Dust Grain Size}",
      journal = {\apj},
         year = 2011,
        month = sep,
       volume = {738},
       number = {2},
          eid = {141},
        pages = {141},
          doi = {10.1088/0004-637X/738/2/141},
archivePrefix = {arXiv},
       eprint = {1106.2682},
 primaryClass = {astro-ph.EP},
       adsurl = {https://ui.adsabs.harvard.edu/abs/2011ApJ...738..141O}
}

@INCOLLECTION{Aikawa2024,
       author = {{Aikawa}, Yuri and {Okuzumi}, Satoshi and {Pontoppidan}, Klaus},
        title = "{The Physical and Chemical Processes in Protoplanetary Disks: Constraints on the Composition of Comets}",
    booktitle = {Comets III},
         year = 2024,
       editor = {{Meech}, Karen. J. and {Combi}, Michael. R. and {Bockel{\'e}e-Morvan}, Dominique and {Raymond}, Sean. N. and {Zolensky}, Michael. E.},
        pages = {33-62},
          doi = {10.2458/azu_uapress_9780816553631-ch002},
       adsurl = {https://ui.adsabs.harvard.edu/abs/2024come.book...33A}
}

@ARTICLE{Banzatti2023,
       author = {{Banzatti}, Andrea and {Pontoppidan}, Klaus M. and {Carr}, John S. and {Jellison}, Evan and {Pascucci}, Ilaria and {Najita}, Joan R. and {Romero-Mirza}, Carlos E. and {{\"O}berg}, Karin I. and {Kalyaan}, Anusha and {Pinilla}, Paola and {Krijt}, Sebastiaan and {Long}, Feng and {Lambrechts}, Michiel and {Rosotti}, Giovanni and {Herczeg}, Gregory J. and {Salyk}, Colette and {Zhang}, Ke and {Bergin}, Edwin A. and {Ballering}, Nicholas P. and {Meyer}, Michael R. and {Bruderer}, Simon and {Jdiscs Collaboration}},
        title = "{JWST Reveals Excess Cool Water near the Snow Line in Compact Disks, Consistent with Pebble Drift}",
      journal = {\apjl},
         year = 2023,
        month = nov,
       volume = {957},
       number = {2},
          eid = {L22},
        pages = {L22},
          doi = {10.3847/2041-8213/acf5ec},
archivePrefix = {arXiv},
       eprint = {2307.03846},
 primaryClass = {astro-ph.EP},
       adsurl = {https://ui.adsabs.harvard.edu/abs/2023ApJ...957L..22B}
}

@ARTICLE{DeSimone2020,
       author = {{De Simone}, Marta and {Ceccarelli}, Cecilia and {Codella}, Claudio and {Svoboda}, Brian E. and {Chandler}, Claire and {Bouvier}, Mathilde and {Yamamoto}, Satoshi and {Sakai}, Nami and {Caselli}, Paola and {Favre}, Cecile and {Loinard}, Laurent and {Lefloch}, Bertrand and {Liu}, Hauyu Baobab and {L{\'o}pez-Sepulcre}, Ana and {Pineda}, Jaime E. and {Taquet}, Vianney and {Testi}, Leonardo},
        title = "{Hot Corinos Chemical Diversity: Myth or Reality?}",
      journal = {\apjl},
         year = 2020,
        month = jun,
       volume = {896},
       number = {1},
          eid = {L3},
        pages = {L3},
          doi = {10.3847/2041-8213/ab8d41},
archivePrefix = {arXiv},
       eprint = {2006.04484},
 primaryClass = {astro-ph.SR},
       adsurl = {https://ui.adsabs.harvard.edu/abs/2020ApJ...896L...3D}
}

@dataset{Tobin_2023_data,
author = {Tobin, John},
publisher = {Harvard Dataverse},
title = {{HDO/H2O Study}},
year = {2023},
version = {V1},
doi = {10.7910/DVN/MDQJEU},
url = {https://doi.org/10.7910/DVN/MDQJEU}
}
\bibliographystyle{aasjournalv7}



\end{document}